# Applying Machine Learning to Elucidate Ultrafast Demagnetization Dynamics in Ni and Ni$_{80}$Fe$_{20}$

Hasan Ahmadian Baghbaderani*, Byoung-Chul Choi

Department of Physics and Astronomy, University of Victoria, Victoria, BC V8W 3P6, Canada

**Abstract**

Understanding the correlation between fast and ultrafast demagnetization processes is crucial for elucidating the microscopic mechanisms underlying ultrafast demagnetization, which is pivotal for various applications in spintronics. Initial theoretical models attempted to establish this correlation but faced challenges due to the complex interplay of physical phenomena. To address this, we employed a variety of machine learning methods, including supervised learning regression algorithms and symbolic regression, to analyze limited experimental data and derive meaningful mathematical expressions between demagnetization time ($\tau_M$) and the Gilbert damping factor ($\alpha$). The results reveal that polynomial regression and K-nearest neighbors algorithms perform best in predicting $\tau_M$. Additionally, sure-independence-screening-and-sparsifying-operator (SISSO) as a symbolic regression method suggested a direct correlation between $\tau_M$ and $\alpha$ for Ni and Ni$_{80}$Fe$_{20}$, indicating spin-flip scattering predominantly influences the ultrafast demagnetization mechanism. The developed models demonstrate promising predictive capabilities, validated against independent experimental data. Comparative analysis between different materials underscores the significant impact of material properties on ultrafast demagnetization behavior. This study underscores the potential of machine learning in unraveling complex physical phenomena and offers valuable insights for future research in ultrafast magnetism.

## 1. Introduction

Combining spin and charge properties, spintronics offers diverse functionalities essential for industrial applications such as sensing and memory storage [1], [2], while also showing promise in communication and information processing [3], [4]. Materials spin textures can be dynamically excited by various physical parameters: magnetic fields, electrical currents, temperature, and pressure, all of which give rise to different responses at different time scales, ranging from nanoseconds (fast demagnetization) to picoseconds (ultrafast demagnetization) [5], [6]. Recent research has intensified the focus on understanding the rapid dynamics of magnetization [7]. However, the mechanisms underlying ultrafast demagnetization dynamics remain poorly understood.

Different theoretical models describe fast and ultrafast magnetization dynamics. For instance, the fast magnetization dynamics can be described through the Landau–Lifshitz–Gilbert (LLG) equation (Eq. 1) [8]:

$$\frac{dM(r,t)}{dt} = -\gamma\big(M(r,t) \times H_{eff}(r,t)\big) + \frac{1}{|M(r,t)|}\big(M(r,t) \times \alpha\frac{dM(r,t)}{dt}\big) \qquad (1)$$

---

* Corresponding author
E-mail address: baghbaderani@uvic.ca

where $M(r,t)$ is the magnetization vector at position $r$ and time $t$, and $\gamma$ is the gyromagnetic ratio that describes the precession of the magnetization around the effective magnetic field, $H_{eff}(r,t)$. This effective field is composed of the external field, the magnetic anisotropy field, and the demagnetization field. The Gilbert damping factor, $\alpha$, quantifies the rate at which the magnetization vector relaxes towards the direction of the effective magnetic field. In other words, the first term on the right-hand side represents precession which includes indirect damping, in which $\gamma$ describes the precession of the magnetization around the micromagnetic effective field, $H_{eff}(r,t)$. The second term in Eq. 1 is the damping term (represented by $\alpha$) which drives the system toward the direction of $H_{eff}$ [9]. On the other hand, ultrafast dynamics following laser excitation are typically modelled using a phenomenological three-temperature model [10], which accounts for electron, spin, and lattice interactions. In this regard, the ultrafast demagnetization time, $\tau_M$, quantifies the rate of ultrafast dynamics. Therefore, $\alpha$ and $\tau_M$ can represent fast and ultrafast demagnetization phenomena, respectively.

Both fast and ultrafast demagnetization, characterized by $\alpha$ and $\tau_M$, require angular momentum transfer from the electronic system to the lattice. Assuming common microscopic channels for angular momentum transfer in both cases, establishing a relation between α and $\tau_M$ becomes desirable [11]. Through several independent investigations relating these two distinct processes [10]–[12], the correlation between $\tau_M$ and α emerged as a critical indicator of the dominant microscopic contribution to the ultrafast demagnetization. It was proposed that $\tau_M$ and α can be either directly or inversely proportional to each other depending on the predominant microscopic mechanism contributing to the magnetic damping [13]:

I) A proportional dependence would suggest a dominating conductivity-like or 'breathing Fermi surface' contribution to the damping due to the scattering of intra-band electrons and holes, where local spin-flip scattering processes are expected to dominate the demagnetization process [14].
II) An inverse dependence may arise in materials with a dominant resistivity-like damping arising from inter-band scattering processes. Zhang et al. [15] suggested that this inverse correlation may arise in the presence of spin transport, which provides an additional relaxation channel resulting in an enhancement of the magnetic damping along with an acceleration of the demagnetization process at the femtosecond timescale.

Based on this, various physical models were proposed to estimate $\tau_M$ through correlating it to α, which results in identifying the microscopic mechanism of ultrafast demagnetization. Initially, efforts were grounded in the assumption that the demagnetization rate correlates with spin-flip processes via the Elliot-Yafet mechanism [12], [16]. Using the LLG equation, an analytical expression was derived, linking $\tau_M$ with α via the Curie temperature, $T_C$ [12]:

$$\tau_M \approx \frac{c_0 \hbar}{k_B T_C \alpha} \qquad (2)$$

with ℏ and $k_B$ the Planck and Boltzmann constants, respectively, and the prefactor $c_0 \sim \frac{1}{4}$. Based on Eq. 2, spin transport can be considered as the dominant mechanism of UFD proved by an inverse correlation between $\tau_M$ and α. This approach aimed to connect the characteristic times of magnon scattering at the center and edge of the Brillouin zone.

Further, the authors of [12]] calculated the transversal spin-relaxation time $\tau_{M, t}$ for laser-excited electrons, describing the damped precessional motion of the single electrons using the LLG equation with **α** and the exchange field, $H_{ex}$. An individual electron spin, which is not aligned with the sea of other electrons, feels this exchange field. Assuming that the longitudinal relaxation time $\tau_M$ is equal to the transverse relaxation time $\tau_{M,t}$ even for the extremely fast precession in the Stoner exchange field yields:

$$\tau_M \approx \frac{\hbar}{g\mu_B H_{ex} \alpha} \tag{3}$$

with the Landé factor g≈ 2 and the Bohr magneton $\mu_B$. Assumptions and simplifications used in driving Eqs. 2 and 3 have cast doubt on the adequacy of the proposed models [17], particularly evident in the models' failure to qualitatively describe experimental results on the demagnetization of rare-earth-doped metals [18]. Based on these models, Eqs. 2 and 3, it is found that $\tau_M$ can be inversely related to α, which describes the damping of GHz precessional motion of the magnetization vector, showcasing spin transport as the dominant mechanism.

In another effort, Fähnle et. al [11] presented a novel derivation of the relationship between α and $\tau_M$. This theory offers a more intricate consideration of a material's electronic properties, resulting in a more complex relationship between α and $\tau_M$, which now involves multiple parameters related to individual electronic states. It is demonstrated that the theory's prediction of α/$\tau_M$ aligns well with experimental observations. The paper elucidates the generation of intra-band and inter-band electron-hole pairs due to changes in spin-orbit energy, contributing to conductivity-like and resistivity-like behaviors, respectively. The study emphasizes the significance of the 'breathing Fermi surface' model. Unlike previous theories, which relied on statistical macrospin approaches, this model adopts a statistical approach at the single-electron level, incorporating detailed dynamics. The derived relation is:

$$\tau_M \approx \frac{M}{\gamma F_{el} pb^2} \alpha \tag{4}$$

where *M* is the material magnetization, $F_{el}$ is a quantity dependent on electronic states' sensitivity to spin-orbit coupling changes, and $pb^2$ represents the degree of spin mixing. Experimental validation, particularly in the case of nickel, confirms the theory's predictive capability. This again establishes a crucial link between magnetization dynamics across different time scales [11], which implies that spin-flip scattering is the dominant mechanism for ultrafast demagnetization, based on the proportional relationship between $\tau_M$ and α (Eq. 4) [14].

The relationship between $\tau_M$ and α presents a complex and largely unresolved challenge, as evidenced by the array of parameters utilized in existing models (Eqs. 2-4) and the varying nature of their relationships. The multitude of factors at play, including material-specific electronic properties, spin-orbit coupling effects, and spin-flip processes, complicates the establishment of a precise correlation between these parameters and the elucidation of the underlying mechanism driving ultrafast demagnetization. The diverse approaches outlined reflect attempts to capture this complexity, yet none offer a definitive solution due to the intricate interplay between different physical phenomena.

Machine learning presents a promising avenue to navigate this complexity, offering the potential to analyze datasets encompassing diverse materials and experimental conditions to discern patterns and relationships that may elude traditional analytical methods. By leveraging machine learning algorithms, it may be possible to develop a predictive model that considers the collective influence of these parameters and sheds light on the dominant mechanisms governing ultrafast demagnetization, facilitating more accurate predictions and deeper insights into this fundamental phenomenon.

In this study, we delve into the intricate relationship between fast and ultrafast demagnetization processes, aiming to elucidate the underlying mechanisms governing magnetization dynamics. Through a comprehensive review of theoretical models, Eqs 1-4, we underscore the challenges in establishing a definitive correlation between $\tau_M$ and $\alpha$ and pinpoint the main descriptors playing roles in such dynamics. Leveraging the power of machine learning techniques, particularly Symbolic Regression (SR) through the sure-independence-screening-and-sparsifying-operator (SISSO) method, we embark on a novel approach to analyze experimental data and unveil meaningful mathematical expressions. Notably, for the first time in this field, machine learning is employed to shed light on the complex interplay of physical phenomena driving ultrafast demagnetization processes. Through the integration of machine learning with experimental data, we aim to provide deeper insights into the fundamental mechanisms of demagnetization, paving the way for advancements in materials science and spintronics applications.

## 2. Methods

As mentioned in the last section, the inherent complexity of ultrafast demagnetization poses challenges for explicit physical modeling, prompting the exploration of alternative approaches such as Machine Learning (ML). In this study, the target property was the UFD time of Ni and $Ni_{80}Fe_{20}$, with a set of 11 input descriptors, all taken from Eqs. 2-4, including $\alpha, T_C, \mu_B, H_{ex}, M, \gamma, F_{el}, pb^2, \hbar, g$ and $k_B$. It is worth mentioning that, as $\alpha$ is a function of various parameters, including experimental conditions such as laser fluence and film thickness, considering $\alpha$ inherently accounts for these experimental conditions. In this regard, a set of data, including 22 and 27 data points for Permalloy and nickel respectively, were collected from [14], [19], [20].

First, a variety of algorithms, including polynomial regression, as well as regression types such as K-nearest neighbors (KNN), decision tree, and support vector machine (SVM), were employed for supervised training. In this regard, the algorithms were trained using 80% of randomly chosen data from both Ni and $Ni_{80}Fe_{20}$, and tested using the remaining 20% of the data. The root-mean-square error (RMSE), Mean Absolute Error (MAE), and R² (R-squared or Coefficient of Determination) were used as performance metrics for the different models utilized.

While ML offers a powerful means to uncover complex relationships, as mentioned conventional methods often demand larger datasets than typically available in materials physics, limiting their applicability. Moreover, such black-box models hinder the elucidation of underlying mechanisms and the identification of key parameters for optimization. One promising avenue bridging physical reasoning with data-centric approaches is Symbolic Regression (SR).

SR seeks to identify interpretable mathematical expressions relating target properties to key input parameters. Within this framework, the sure-independence-screening-and-sparsifying-operator (SISSO) approach emerges as a novel methodology for identifying analytical expressions, leveraging compressed sensing principles to navigate the vast solution space efficiently [21]. SISSO methodology begins by assembling a set of primary descriptors relevant to the target property. These descriptors are then expanded into a vast pool of potential expressions through the iterative application of various mathematical operators. Compressed sensing techniques are employed to identify the optimal descriptor vector, which contains the most relevant expressions, thus enabling the creation of low-dimensional models. Through the initial descriptor selection using Sure Independence Screening (SIS) and subsequent recursive optimization with Sparsifying Operators (SO), SISSO reselects descriptors and reoptimizes coefficients at each step. This dynamic approach balances accuracy and computational efficiency, overcoming the limitations of traditional sparse solution algorithms and allowing for the identification of the best sparse solution from high-dimensional data [22]. However, the combinatorial explosion of the solution space presents computational challenges, underscoring the need for effective variable selection methods to navigate this complexity.

Variable selection techniques, including filter, wrapper, and embedded methods, play a pivotal role in refining models and extracting meaningful information from the vast pool of input descriptors. These methods, integrated within the SISSO framework as VS-SISSO, hold promise for improving the learning and understanding of material properties, paving the way for advancements in UFD research and materials design [23].

Symbolic regression, empowered by VS-SISSO, offers a data-driven approach capable of uncovering meaningful insights from limited datasets [23]–[25], as encountered in the study of ultrafast demagnetization. By intelligently exploring the solution space and prioritizing interpretability, SR facilitates the discovery of mathematical expressions that not only accurately capture complex relationships but also offer physical insights into underlying processes. Unlike conventional ML methods, SR produces human-readable formulas, enhancing our understanding of the phenomena under investigation.

All the data from Ni and $Ni_{80}Fe_{20}$ were used to drive the optimal equation to calculate the UFD time. The VS-SISSO algorithm employed two operators, division and multiplication, and the maximal descriptor complexity for all jobs was set to 7, representing the maximum number of operators used in the learning process to derive each equation. The descriptor dimension was set to one, indicating the number of terms in the equation, and the intercept was set to zero. These parameters were selected based on the rationale that we aimed to derive an equation that provides a better understanding of the ultrafast demagnetization mechanism, predicts the results more accurately, and is consistent with the units of ultrafast demagnetization time. Then, the model was tested using new set of experimental data.

## 2. Results and Discussions

### 2.1 Supervised Learning Methods

A variety of supervised learning methods were employed to predict the ultrafast demagnetization (UFD) time based on selected descriptors. In this context, polynomial regression, KNN, decision tree, and SVM algorithms were trained and tested. During the training process, utilizing 80% of the data, the parameters of these algorithms were optimized to achieve the highest accuracy. As a case in point, among polynomial regression algorithms with degrees in the ranges of 1-10, polynomial regression with a second degree offers the lowest error. Figure 1 illustrates the parity plots for the different ML algorithms. To better understand the prediction errors associated with these methods, Table 1 lists the values for various error metrics, including RMSE, MAE, and R². As can be seen from the figures and the table, polynomial regression and KNN provide the best performance. Although these algorithms can predict the $\tau_M$ for new samples using the selected descriptors, they do not elucidate the underlying mechanisms of UFD. Therefore, symbolic regression, using SISSO, was employed to gain a better understanding of the relationship between $\tau_M$ and α. This can be used to identify the dominant mechanism of UFD.

Table 1. Error metrics of different supervised algorithms trained by Ni and $Ni_{80}Fe_{20}$ data.

| Supervised Learning Methods | Error Metrics | | |
|---|---|---|---|
| | **RMSE** | **MAE** | $R^2$ |
| Polynomial Regression | 1.0603 | 0.8453 | 0.9996 |
| KNN | 5.5292 | 4.1799 | 0.9894 |
| Decision Tree | 9.0111 | 6.6000 | 0.9718 |
| SVM | 19.2800 | 13.220 | 0.8710 |

### 2.2 Symbolic Regression Method

The data of two materials were used to train two individual VS−SISSO algorithm to find the most accurate equation explaining the UFD procedure. It was found that for both Ni and $Ni_{80}Fe_{20}$, there is a direct correlation between $\tau_M$ and $\alpha$, so for both cases by feeding all the available descriptors into VS−SISSO, ML leads to the optimal equation of:

$$\tau_M = c_1 \alpha \qquad (5)$$

$c_1$ as a constant number is different for Ni and Permalloy. Equation 5 shows that VS-SISSO utilizes only the values of α from the 22 and 27 samples of Permalloy and Nickel to generate the most accurate formula for each case individually. It disregards the remaining parameters, which are constant for all considered samples in each material. This is because VS−SISSO integrates symbolic regression (SISSO) with iterative variable selection (VS), automatically searching for important variables from a pool of input descriptors [23]. A direct correlation between $\tau_M$ and α indicates that spin-flip scattering can be the dominant microscopic mechanism for UFD and a major intra-band conductivity-like contribution to α. The comparison between the experimental and predicted results can be seen in Figure 2. Based on the error metrics mentioned in Table 2, there is a better fit between the experimental and modelled $\tau_M$ in the case of Ni.

After making sure that the dominant UFD mechanism in both materials is the same, the whole set of data and descriptors for both Ni and Ni$_{80}$Fe$_{20}$ were introduced to VS−SISSO and the following optimal equation was suggested:

$$\tau_M = c_1 \frac{T_C^6 g^3}{\gamma M} \alpha \tag{6}$$

The parity plot and the details of the errors, e.g. RMSE= 12.68, between experimental and modelled $\tau_M$ can be seen in Fig. 3a and Table 2, respectively. In order to make the unit of the output model right a correction is done and the suggested model is:

$$\tau_M = c_1 \frac{T_C^6 g^3}{T_D^6 \gamma M} \alpha \tag{7}$$

where $T_D$ is Debye temperature. Although the RMSE= 17.36 of Eq. 7 is not as good as the one for the last suggested model, Eq. 6, as it can be seen from Fig. 3b, there is a fairly good agreement between the experimental and calculated demagnetization time based on the suggested model. It's worth noting that, unlike many models proposed by symbolic regression, where the resulting units may not align correctly, the output unit of this model is consistent with the predicted parameter, $\tau_M$. Both are in femtoseconds, ensuring coherence in the results. This general relationship also suggests that $\tau_M \alpha \alpha$ and as a result the spin-flip scattering is the dominant mechanism of UFD.

The relationship between $\tau_M$ and different effective parameters, mentioned in the model (Eq. 7), can be tested based on the independent data which were not used to obtain the current models. In accordance with the model proposed by [7], it is noted that the $\tau_M$ of Permalloy surpasses that of Ni. This observation is consistent with the output of our proposed model (Eq. 7), where a substantial disparity in $T_C$ is evident in favor of Permalloy, outweighing the marginal variation in $T_D$ favoring Ni. It is noteworthy that the higher magnetic moment of Permalloy fails to fully offset the considerable impact of Curie temperature, given its exponent of six. This trend aligns with the comparison of experimental findings in [7] for Ni and Permalloy. This evaluation confirms the validity of the achieved model in predicting experimental and modelling test trends.

In addition, the model was used to predict the $\tau_M$ of new Permalloy and Ni samples. In this regard, in the work done in [14], it was found that in Ni$_{80}$Fe$_{20}$ samples, when α≃ 0.012, the $\tau_M$ measured to be ≃ 225 fs and our model suggests $\tau_M$≃ 265 fs. As it can be seen, there is a fairly good agreement between the experimental and predicted data by the model, Eq. 7. In addition, De et al. [20] experimentally investigated the coherent spin dynamics after ultrashort laser excitation by performing time-resolved magneto-optical Kerr effect experiments on Ni$_{80}$Fe$_{20}$ thin films. They reported $\tau_M$≃ 200 fs when α≃ 0.00901. Based on this data, our proposed model, Eq. 7, gives rise to $\tau_M$≃ 198.8 fs which is in very good agreement with the experimental results. Koopmans et al. [12] chose an all-optical approach in which both demagnetization and precession can be triggered by a single laser pulse and successively probed the demagnetization time ($\tau_M$≃ 150 fs), when α≃ 0.038. Using this data, our model suggests $\tau_M \simeq 213$ fs. Therefore, it seems that there is a fairly good agreement between experimental and modelled results both for Ni and Ni$_{80}$Fe$_{20}$.

Table 2 Error metrics of VS-SISSO algorithms trained by Ni and Ni$_{80}$Fe$_{20}$ (Py) data.

| Symbolic Regression Method | Training Data | Error Metrics | | |
|---|---|---|---|---|
| | | RMSE | MAE | $R^2$ |
| VS-SISSO | Ni | 1.6404 | 1.1787 | 0.9873 |
| VS-SISSO | Py | 11.0814 | 9.4609 | 0.9543 |
| VS-SISSO | Ni, Py | 12.6785 | 11.5872 | 0.9577 |
| VS-SISSO (Unit Corrected) | Ni, Py | 17.3552 | 15.6498 | 0.9208 |

## Conclusions

In conclusion, the study integrates a machine learning technique with experimental data to explore the ultrafast demagnetization behavior of Ni and Ni$_{80}$Fe$_{20}$ alloys. Among algorithms used for supervised learning, second-degree polynomial regression and KNN offer the lowest error. Through the VS−SISSO method, a direct correlation between demagnetization time ($\tau_M$) and Gilbert damping parameter ($\alpha$) is established for both materials, indicating the significance of spin-flip scattering in the demagnetization mechanism. The developed models demonstrate promising predictive capabilities, with favorable agreement between predicted and observed $\tau_M$ values for both alloys. The consistency in output units ensures coherence in results, while validation against independent experimental data further supports the models' robustness. Moreover, comparisons between Ni and Permalloy highlight the models' ability to capture material property influences on ultrafast demagnetization behavior. Overall, this study showcases the efficacy of machine learning in elucidating complex phenomena and offers valuable insights for future experimental investigations and technological advancements in ultrafast magnetism.

## References

[1] R. Ismael Salinas, P. C. Chen, C. Y. Yang, and C. H. Lai, "Spintronic materials and devices towards an artificial neural network: accomplishments and the last mile," *Mater. Res. Lett.*, vol. 11, no. 5, pp. 305–326, 2023, doi: 10.1080/21663831.2022.2147803.

[2] A. Hirohata *et al.*, "Review on spintronics: Principles and device applications," *J. Magn. Magn. Mater.*, vol. 509, no. March, 2020, doi: 10.1016/j.jmmm.2020.166711.

[3] J. Grollier, D. Querlioz, K. Y. Camsari, K. Everschor-Sitte, S. Fukami, and M. D. Stiles, "Neuromorphic spintronics," *Nat. Electron.*, vol. 3, no. 7, pp. 360–370, 2020, doi: 10.1038/s41928-019-0360-9.

[4] N. Leroux *et al.*, "Radio-Frequency Multiply-and-Accumulate Operations with Spintronic Synapses," *Phys. Rev. Appl.*, vol. 15, no. 3, pp. 1–11, 2021, doi: 10.1103/PhysRevApplied.15.034067.

[5] X. Chen *et al.*, "Forecasting the outcome of spintronic experiments with Neural Ordinary Differential Equations," *Nat. Commun.*, vol. 13, no. 1, pp. 1–12, 2022, doi: 10.1038/s41467-022-28571-7.

[6] E. Beaurepaire, J. C. Merle, A. Daunois, and J. Y. Bigot, "Ultrafast spin dynamics in ferromagnetic nickel," *Phys. Rev. Lett.*, vol. 76, no. 22, pp. 4250–4253, 1996, doi: 10.1103/PhysRevLett.76.4250.

[7] S. M. Hosseini, F. Jahangiri, R. Jalilian, and S. M. Hamidi, "Comparative study of femtosecond laser-induced ultrafast magnetization dynamics in soft ferromagnetic ultra-thin alloy," *J. Magn. Magn. Mater.*, vol. 579, 2023, doi: 10.1016/j.jmmm.2023.170879.

[8] T. L. Gilbert, "A phenomenological theory of damping in ferromagnetic materials," *IEEE Trans. Magn.*, vol. 40, no. 6, pp. 3443–3449, 2004, doi: 10.1109/TMAG.2004.836740.

[9] M. Fähnle and C. Illg, "Electron theory of fast and ultrafast dissipative magnetization dynamics," *J. Phys. Condens. Matter*, vol. 23, no. 49, 2011, doi: 10.1088/0953-8984/23/49/493201.

[10] B. Koopmans *et al.*, "Explaining the paradoxical diversity of ultrafast laser-induced demagnetization," *Nat. Mater.*, vol. 9, no. 3, pp. 259–265, 2010, doi: 10.1038/nmat2593.

[11] M. Fähnle, J. Seib, and C. Illg, "Relating Gilbert damping and ultrafast laser-induced demagnetization," *Phys. Rev. B - Condens. Matter Mater. Phys.*, vol. 82, no. 14, pp. 1–4, 2010, doi: 10.1103/PhysRevB.82.144405.

[12] B. Koopmans, J. J. M. Ruigrok, F. Dalla Longa, and W. J. M. De Jonge, "Unifying ultrafast magnetization dynamics," *Phys. Rev. Lett.*, vol. 95, no. 26, pp. 1–4, 2005, doi: 10.1103/PhysRevLett.95.267207.

[13] W. Zhang, W. He, X. Q. Zhang, Z. H. Cheng, J. Teng, and M. Fähnle, "Unifying ultrafast demagnetization and intrinsic Gilbert damping in Co/Ni bilayers with electronic relaxation near the Fermi surface," *Phys. Rev. B*, vol. 96, no. 22, pp. 1–7, 2017, doi: 10.1103/PhysRevB.96.220415.

[14] S. Mukhopadhyay, S. Majumder, S. Narayan Panda, and A. Barman, "Investigation of ultrafast demagnetization and Gilbert damping and their correlation in different ferromagnetic thin films grown under identical conditions," *Nanotechnology*, vol. 34, no. 23, 2023, doi: 10.1088/1361-6528/acc079.

[15] W. Zhang *et al.*, "Enhancement of ultrafast demagnetization rate and Gilbert damping driven by femtosecond laser-induced spin currents in F e81 G a19/ i r20 M n80 bilayers," *Phys. Rev. B*, vol. 100, no. 10, pp. 1–11, 2019, doi: 10.1103/PhysRevB.100.104412.

[16] B. Koopmans, H. H. J. E. Kicken, M. Van Kampen, and W. J. M. De Jonge, "Microscopic model for femtosecond magnetization dynamics," *J. Magn. Magn. Mater.*, vol. 286, no. SPEC. ISS., pp. 271–275, 2005, doi: 10.1016/j.jmmm.2004.09.079.

[17] J. Walowski *et al.*, "Energy equilibration processes of electrons, magnons, and phonons at the femtosecond time scale," *Phys. Rev. Lett.*, vol. 101, no. 23, pp. 1–4, 2008, doi: 10.1103/PhysRevLett.101.237401.

[18]  I. Radu *et al.*, "Laser-induced magnetization dynamics of lanthanide-doped permalloy thin films," *Phys. Rev. Lett.*, vol. 102, no. 11, pp. 1–4, 2009, doi: 10.1103/PhysRevLett.102.117201.

[19]  S. N. Panda, S. Mondal, S. Majumder, and A. Barman, "Ultrafast demagnetization and precession in permalloy films with varying thickness," *Phys. Rev. B*, vol. 108, no. 14, pp. 1–12, 2023, doi: 10.1103/PhysRevB.108.144421.

[20]  A. De *et al.*, "Coherent and incoherent magnons induced by strong ultrafast demagnetization in thin permalloy films," vol. 024422, pp. 1–7, 2023, doi: 10.1103/PhysRevB.109.024422.

[21]  R. Ouyang, E. Ahmetcik, C. Carbogno, M. Scheffler, and L. M. Ghiringhelli, "Simultaneous learning of several materials properties from incomplete databases with multi-task SISSO," *JPhys Mater.*, vol. 2, no. 2, 2019, doi: 10.1088/2515-7639/ab077b.

[22]  R. Ouyang, S. Curtarolo, E. Ahmetcik, M. Scheffler, and L. M. Ghiringhelli, "SISSO: A compressed-sensing method for identifying the best low-dimensional descriptor in an immensity of offered candidates," *Phys. Rev. Mater.*, vol. 2, no. 8, pp. 1–11, 2018, doi: 10.1103/PhysRevMaterials.2.083802.

[23]  Z. Guo, S. Hu, Z. K. Han, and R. Ouyang, "Improving Symbolic Regression for Predicting Materials Properties with Iterative Variable Selection," *J. Chem. Theory Comput.*, vol. 18, no. 8, pp. 4945–4951, 2022, doi: 10.1021/acs.jctc.2c00281.

[24]  C. J. Bartel *et al.*, "New tolerance factor to predict the stability of perovskite oxides and halides," *Sci. Adv.*, vol. 5, no. 2, pp. 1–10, 2019, doi: 10.1126/sciadv.aav0693.

[25]  S. R. Xie, G. R. Stewart, J. J. Hamlin, P. J. Hirschfeld, and R. G. Hennig, "Functional form of the superconducting critical temperature from machine learning," *Phys. Rev. B*, vol. 100, no. 17, pp. 1–6, 2019, doi: 10.1103/PhysRevB.100.174513.

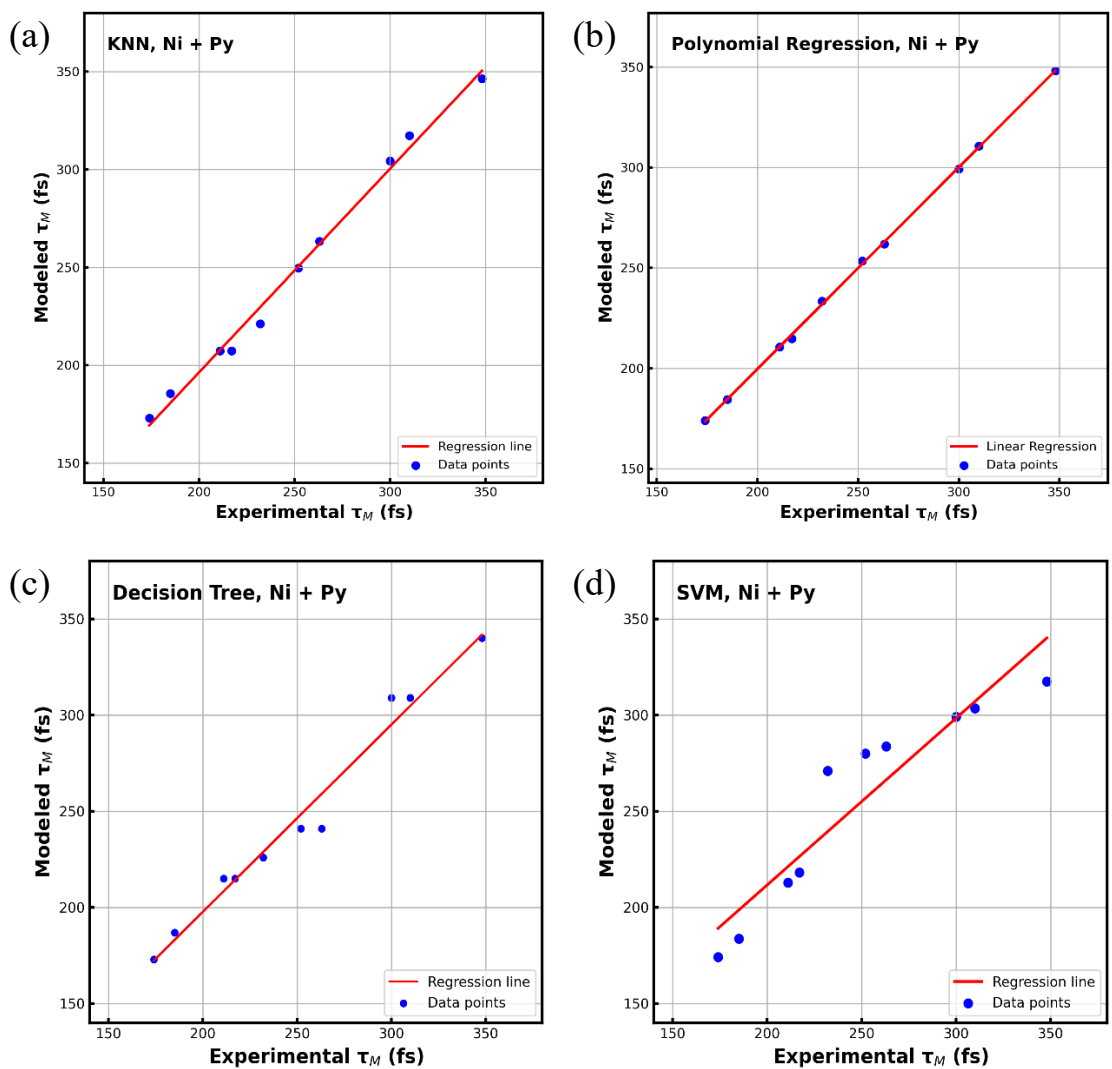

Figure 1. Parity plots of supervised ML methods, (a) Polynomial Regression, (b) KNN, (c) Decision Trees and (d) SVM) to compare the experimental vs modelled predicted ultrafast demagnetization times of test data of Ni and $Ni_{80}Fe_{20}$ (Py).

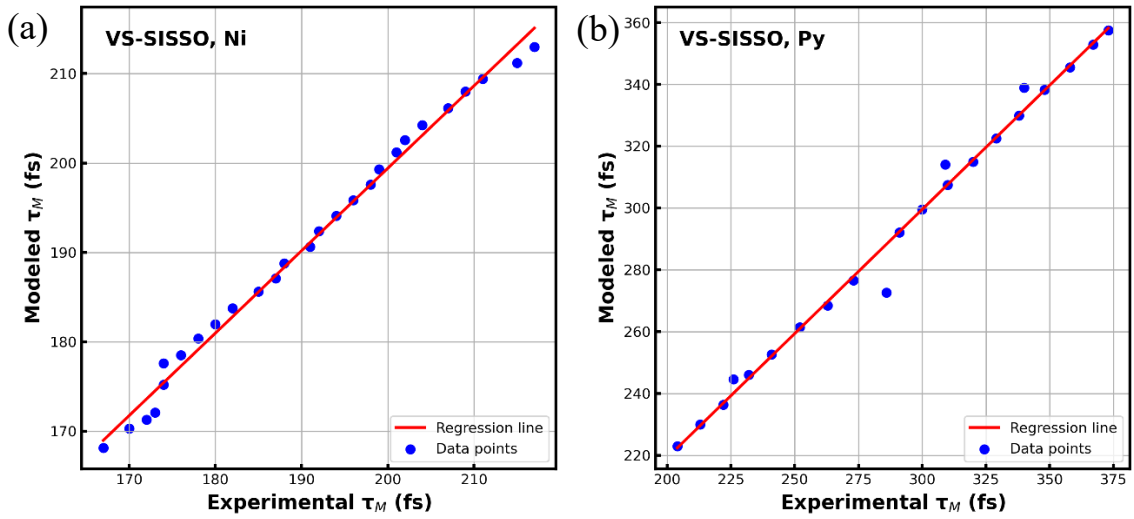

Figure 2. Parity plots of VS-SISSO algorithm to get analytical expression for $\tau_M$ as a function of all the available features for (a) Ni and (b) $Ni_{80}Fe_{20}$ (Py) samples, when 2 operators were considered to build the model.

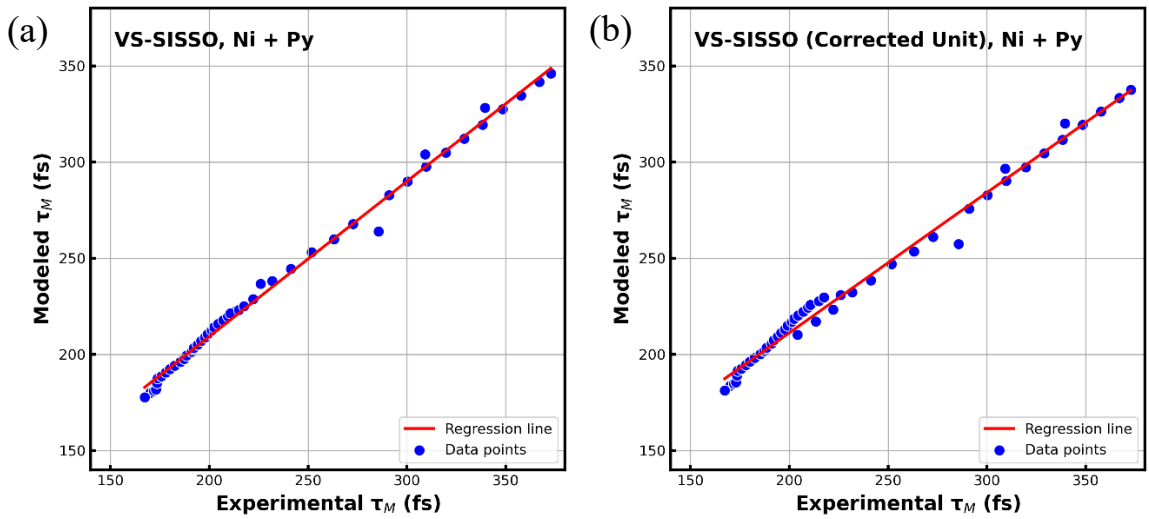

Figure 3. Parity plots of (a) VS-SISSO algorithm to get analytical expression for $\tau_M$ as a function of all the available features for Ni and $Ni_{80}Fe_{20}$ (Py) samples, (b) corrected VS-SISSO driven relationship based on unit conversion.